\documentclass[aps,prl,twocolumn,superscriptaddress,nofootinbib]{revtex4-1}
\usepackage[utf8]{inputenc}
\usepackage{graphicx}
\usepackage{xcolor}
\usepackage{amsmath}
\usepackage{ulem}
\begin{document}
\title{
Microscopic foundation of the $\mu$(I) rheology for dense granular flows on inclined planes.
}
\author{Denis Dumont}
\affiliation{Laboratoire Interfaces $\&$ Fluides Complexes, Universit\'e de Mons, 20 Place du Parc, B-7000 Mons, Belgium.}
\author{Haggai Bonneau}
\affiliation{UMR CNRS Gulliver 7083, ESPCI Paris, PSL Research University, 75005 Paris, France.}
\author{Thomas Salez}
\affiliation{Univ. Bordeaux, CNRS, LOMA, UMR 5798, F-33400 Talence, France.}
\author{Elie Raphael}
\affiliation{UMR CNRS Gulliver 7083, ESPCI Paris, PSL Research University, 75005 Paris, France.}
\author{Pascal Damman}
\email{pascal.damman@umons.ac.be}
\affiliation{Laboratoire Interfaces $\&$ Fluides Complexes, Universit\'e de Mons, 20 Place du Parc, B-7000 Mons, Belgium.}
\date{\today}

\begin{abstract}
Macroscopic and microscopic properties of dense granular layers flowing down inclined planes are obtained from Discrete-Element-Method simulations for both frictionless and frictional grains.  Three fundamental observations for dense granular flows are recovered, namely the occurrence of a critical stress, the Bagnold velocity profile, as well as well-defined friction and dilatancy laws. The microscopic aspects of the grain motion highlight the formation of transient clusters. From this microscopic picture, we derive a theoretical scaling model without any empirical input that explains quantitatively the fundamental laws of dense granular flows in incline plane and shear geometries. The adequacy between the model and the observed results suggests that granular flows can be viewed as flows from thermal fluids of hard spheres.
\end{abstract}
\maketitle

Despite several decades of intense research, the mechanisms underlying dense granular flows remain largely misunderstood. A universal framework allowing one to describe the numerous configurations and observations studied in the laboratory is still lacking~\cite{GDRMiDi2004}. Most models remain semi-empirical and are not supported by strong microscopic justifications~\cite{Jop2006,BCRE1994,Bouzid2015,Kamrin2019}. The global flow properties are usually described using the popular $\mu(I)$ rheology. This approach consists in two empirical relations between the macroscopic friction coefficient $\mu$ (defined as the ratio between the shear stress and the pressure) or the volume fraction $\phi$ on one hand, and the inertial number $I=\dot{\gamma} d \sqrt{\rho_\textrm{p}/P}$ on the other hand, involving the shear rate $\dot{\gamma}$, the grain size $d$, the mass density $\rho_\textrm{p}\sim m/d^3$ of the grains, their individual mass $m$, and the pressure $P$~\cite{Jop2006,Andreotti2013}. Essentially, in this Amontons-Coulomb-like description, a granular layer starts to flow when the applied shear stress overcomes a critical frictional stress $\mu_\textrm{c} P$. Nevertheless, this description fails to properly rationalize some important observable features, such as the presence of a metastable region~\cite{Silbert2003,Forterre2003} and the layer-thickness dependence of the angle at which the flow stops~\cite{Pouliquen1999,Borzsonyi2008,Malloggi2015,Perrin2021}. These last decades, it has been shown that nonlocal/cooperative effects are mandatory to properly describe dense granular flows~\cite{Bouzid2013,Kamrin2012,Goyon2008,Henann2013,Tang2018,Saitoh2019,Kamrin2015,DumontSoulard2020}.

In this Letter, using a combination of Discrete-Element-Method (DEM) simulations and a model based on microscopic arguments, we address the rheology of dense granular matter from the canonical setting of a layer flowing down an inclined plane. Therein, the inclination angle $\theta$ and the layer thickness $H$ are the two external control parameters. Previous experimental and numerical studies have shown that the local average velocity profile of a thick granular layer flowing over an inclined plane exhibits a so-called Bagnold profile~\cite{GDRMiDi2004,Silbert2001,Baran2006}, \textit{i.e.} $\langle v(z,t)-v(0,t)\rangle \sim H^{3/2}-(H-z)^{3/2}$, where $v(z,t)$ is the local velocity field along the flow direction, at normal coordinate $z$ and time $t$. Besides, it has been suggested that nonlocal cooperative effects are essential to describe the layer-thickness dependance of the stop angle~\cite{Kamrin2015,DumontSoulard2020}, \textit{i.e.} the smallest angle for which a stationary flow is observed. 
We will see here that the mechanical noise related to grain-grain collisions determines an effective temperature. This concept coupled to the formation of clusters appears to be a fundamental issue to derive a model for granular flow based on the hard sphere fluid limit. The proposed model is able to predict the size of dynamic clusters, the Bagnold velocity profile as well as the two empirical relations, $\mu(I)$ and $\phi(I)$, commonly used to fit experimental and numerical data~\cite{Jop2006,Andreotti2013}. 

\begin{figure}
\begin{center} 
\includegraphics[width=1\linewidth]{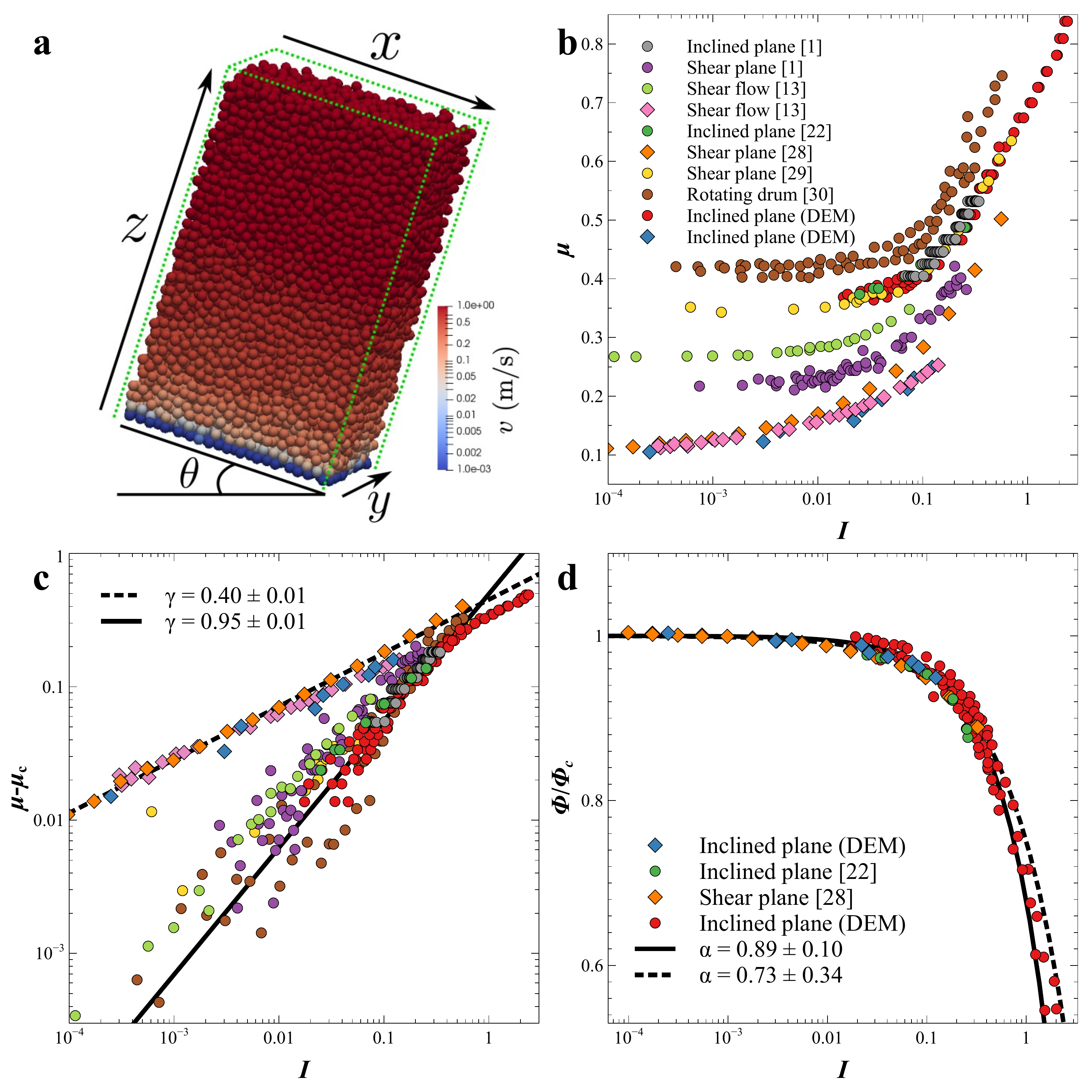}
\caption{a) Typical snapshot of a DEM simulation, with initial layer thickness $H_{\textrm{i}}=30\,d$ and inclination angle $\theta=24^\circ$. The color code indicates the velocity $v(z,t)$. 
b) Macroscopic friction coefficient $\mu$ as a function of inertial number $I=\dot{\gamma}d \sqrt{\rho_{\textrm{p}}/P}$ for frictionless (diamonds) and frictional (circles) grains, as well as various inclination angles $\theta$, initial layer thicknesses $H_{\textrm{i}}$, and various setup configurations~\cite{GDRMiDi2004,Baran2006,Peyneau2008,Bouzid2013,Azema2014,Li2020}. 
c) Difference $\mu- \mu_\textrm{c}$ in friction coefficient as a function of inertial number $I$, where $\mu_\textrm{c}=\mu(I\rightarrow 0)$, for the same data as in the previous panel. The solid and dashed lines correspond to fit with $\mu - \mu_\mathrm{c} \sim I^{\gamma}$, the values of $\gamma$ are provided in legend.
d) Ratio of volume fraction $\phi/\phi_\mathrm{c}$ as a function of inertial number $I$ for frictionless (diamonds, $\phi_\textrm{c}\simeq 0.64$) and frictional (circles, $\phi_c\simeq 0.6$) grains. Data from~\cite{Baran2006,Peyneau2008} are added for comparison. The solid and dashed lines correspond to fit with $\phi/\phi_c=1 - a I^{\alpha}$, the values of $\alpha$ are provided in legend. }
\label{Fig1}
\end{center}
\end{figure}

The numerical simulations were performed with the software LIGGGHTS~\cite{Kloss2012}. The system consists in a layer of identical grains, with diameter $d=1$~mm, mass $m=\frac{4}{3} \pi \rho_\textrm{p} d^3/8$ and elastic modulus, $E=1 \text{MPa}$, placed on an inclined plane with an inclination angle $\theta$ (see Fig.~\ref{Fig1}a). We focus here on thick-enough layers, in order to avoid the thickness dependence of the stop angle observed for thin layers~\cite{Pouliquen1999,Borzsonyi2008,Malloggi2015,Perrin2021}. The mechanical properties of the simulated grains are set to be exactly the same as in our previous study~\cite{DumontSoulard2020}, and correspond to glass beads~\cite{Pouliquen1999}. In particular, the microscopic coefficients $\mu_\textrm{s}$ and $\mu_\textrm{r}$ of sliding and rolling frictions are set to 0.5 and 0.01, respectively. In addition, frictionless grains (\textit{i.e.}, $\mu_\textrm{s}$=$\mu_\textrm{r}$=0) are also simulated.  The influence of the microscopic friction has also been studied, see SI~\cite{supmat}. The substrate is made of immobile grains to mimic the glued grains in inclined-plane experiments. We impose periodic boundary conditions in the $x$ and $y$ directions to get rid of side-wall effects~\cite{Jop2005}. The size of the base has been carefully chosen in order to be large enough to avoid autocorrelations due to periodicity. We stress that similar set-ups have already been reported~\cite{Silbert2001,Silbert2003,Baran2006}. 

Before the  inclination of the plane, the layer has an initial vertical thickness $H_\textrm{i}$ ranging between $10\,d$ and $60\,d$, with a base of $20\,d\times 20\,d$ in the horizontal plane. The plane is subsequently inclined briefly at $30^{\circ}$ to initiate the flow. Subsequently, the inclination is fixed at the desired angle $\theta$, ranging between $20^{\circ}$ and $40^{\circ}$. For each value of $H_\textrm{i}$ and $\theta$, the actual layer thickness $H$ along $z$, and the mean volume fraction $\phi$ of the whole layer (averaged over at least 10 time steps in the steady state) are measured. The average local velocity profiles $\langle v(z,t)\rangle$ and the inertial number $I$ are also computed. 
As a remark, we have the relation $\dot{\gamma}(z)=\textrm{d}\langle v(z,t)\rangle/\textrm{d}z$. The averages $\langle\rangle$ are performed over time and realizations, at fixed $z$.

In agreement with previous works~\cite{GDRMiDi2004,Silbert2001,Silbert2003,Baran2006}, we observe that: i) there is a critical stress to induce flow for dense granular layers, corresponding to a macroscopic friction coefficient $0.2\le \mu_\textrm{c} \le 0.4$ for frictional grains, and $0.1$ for frictionless grains (Fig.~\ref{Fig1}b); ii) the local average velocity profile is well described by a Bagnold profile (see Fig.~S2a in SI~\cite{supmat}); iii) the volume fraction $\phi$ (see Fig.~S2b in~\cite{supmat}) and the inertial number $I$ (see Fig.~S2c in SI~\cite{supmat}) remain mostly constant throughout the layer, for all the studied inclination angles. 
As proposed in several studies~\cite{daCruz2005,lois2005,Andreotti2013}, dimensional analysis shows that only one dimensionless parameter is required to describe granular flows, {\it i.e.}, the inertial number $I$ (besides the microscopic friction coefficient). The flow properties are characterized through the frictional, $\mu=\mu(I)$, and the dilatancy, $\phi=\phi(I)$ laws. The macroscopic friction coefficient $\mu$ is determined by the shear to normal stress ratio~\cite{daCruz2005,Jop2006,Andreotti2013}.
For the inclined-plane geometry considered here, both the macroscopic friction coefficient and the pressure are prescribed through the inclination $\theta$ of the plane and the height $H$ of the flowing layer~\cite{daCruz2005}.
In a continuum-limit approximation, the effective friction coefficient for this setup is thus fixed to a constant value, $\mu=\tan(\theta) \simeq\theta$ for the range of inclination angles of interest. From dimensional analysis and since $\mu$ does not depend on $z/d$, we can conclude that $I$ and $\phi$ are constant throughout the layer and fully determined by the inclination angle $\theta$ and the microscopic friction coefficient.

As previously shown, Fig.~\ref{Fig1}c confirms that $\mu(I)$ is well described by $\mu-\mu_\textrm{c}\sim I^\gamma$, with $\gamma =0.40\pm 0.01$ for frictionless grains~\cite{Peyneau2008,Bouzid2013}. For frictional grains, $\gamma = 0.95\pm0.01$  for moderate inclination angles ($I\lesssim 0.1$) in agreement with previous observations ~\cite{GDRMiDi2004,Baran2006,Azema2014,Li2020}. It should be noted that for large inclination angles, we observe a change of the exponent that becomes close to the value of frictionless systems $\gamma =0.4$. 
The exponent for frictional grains does not depend on the (finite) values of the microscopic friction coefficients (see Fig.~S1a in~\cite{supmat}), thus indicating the singularity of the frictionless limit. In contrast, $\mu_\textrm{c}$ depends on the microscopic friction coefficients, but even for frictionless assemblies a non-zero value close to 0.1 is observed~\cite{Peyneau2008,supmat,Perrin2021}. The exact origin of this residual macroscopic friction remains unclear, but should be related to the steric contraints associated with granular topography~\cite{Peyneau2008b}.  

The dilatancy laws obtained from the DEM simulations are shown in Fig.~\ref{Fig1}d and compared to data from the literature~\cite{daCruz2005,Pouliquen2006,Forterre2008}.  For all these combined data, the evolution of the packing fraction with $I$ can be empirically described by the relation $\phi_\textrm{c} - \phi \sim I^\alpha$, where $\phi_\textrm{c}=\phi(I\rightarrow0)$ is the volume fraction at kinetic arrest, and with $\alpha=0.89\pm 0.1$ and $0.73\pm 0.34$, for frictional and frictionless grains, respectively. Note that for frictionless grains,  another functional form was proposed~\cite{Peyneau2008}: $1/\phi - 1/\phi_\textrm{c} \sim I^{0.4}$ but remains valid only for very small inertial numbers, {\it i.e.} $I\lesssim 10^{-2}$.

Hereafter, we investigate the microscopic origin of these laws. As proposed by several authors, the velocity fluctuations and the diffusion coefficient of the grains are strong indicators of their dynamics~\cite{Kharel2017,Saitoh2020}. 
\begin{figure}
\begin{center}
\includegraphics[width=\linewidth]{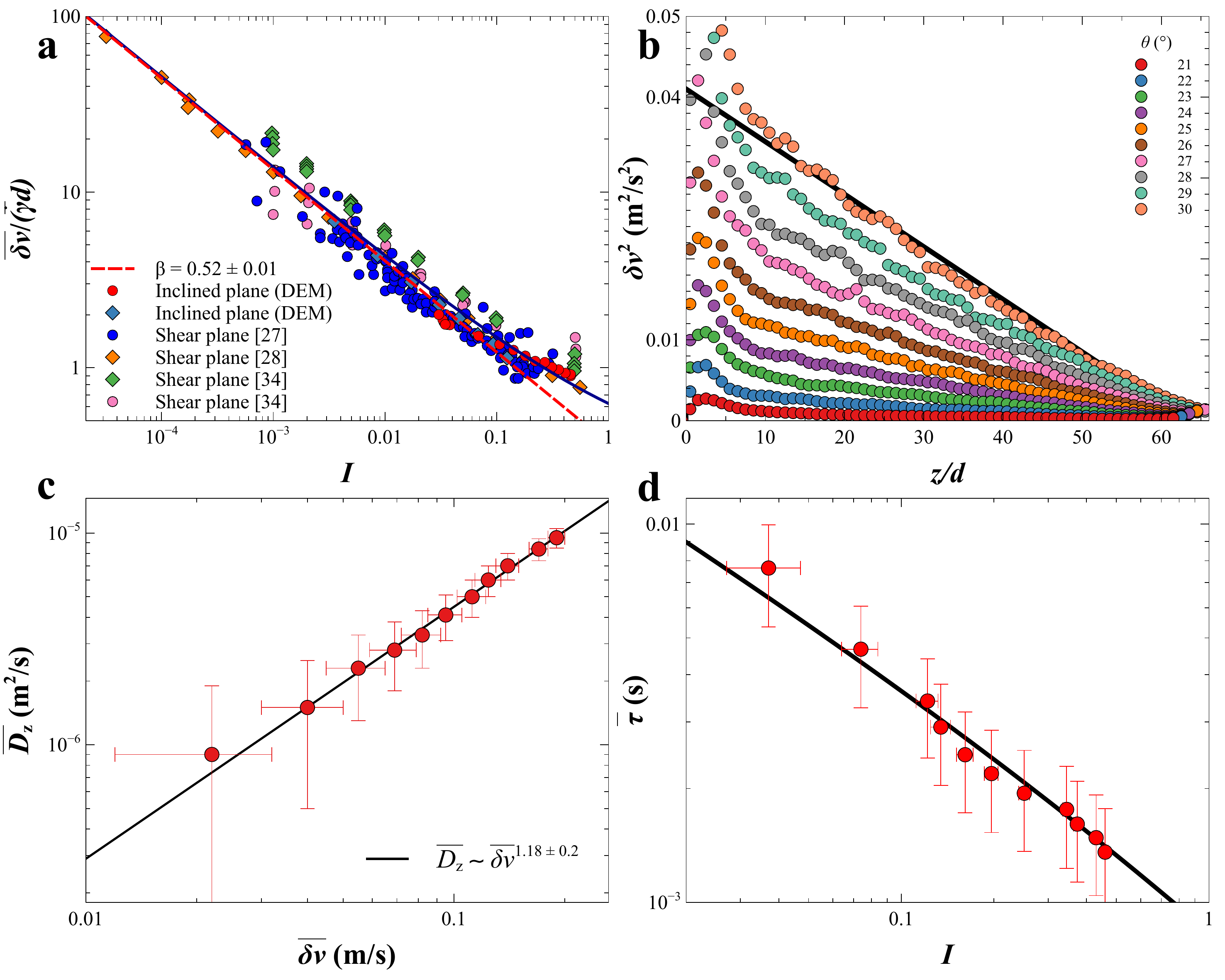}
\caption{a) Thickness-averaged standard deviation $\overline{\delta v}$ of the velocity field normalized by a typical shear velocity $\overline{\dot{\gamma}}d$, as a function of the inertial number $I$, for frictionless (diamonds) and frictional (circles) grains. Results from previous works~\cite{daCruz2005,Peyneau2008,Kharel2017} are also shown for comparison. The dashed line correspond to fit with $\delta v/\dot{\gamma} \sim I^{-\beta}$, the value of $\beta$ is provided in legend (the fit is limited to small $I$'s values, $I < 0.05$). The solid line is a guide for the eyes and corresponds to $a (1 + b/I^{0.52})$ with $a$=0.25 and $b$=1.5.
b) Local variance $\delta v^2(z)= \langle|v(z,t)-\langle v(z,t)\rangle|^2\rangle$ of the velocity field $v(z,t)$, as a function of rescaled normal coordinate $z/d$, for a layer of frictional grains initially characterized by $H_{\textrm{i}}=60\,d$ and various inclination angles $\theta$ as indicated. An affine solid line is added as a guide for the eye. 
c) Thickness-averaged diffusion coefficient $\overline{D_z}$ along $z$ as a function of the thickness-averaged standard deviation $\overline{\delta v}$ of the velocity field for fictional grains. The solid line indicates a fitted expression as provided in the legend.
d) Correlation time $\overline{\tau}$ of the thickness-averaged velocity fluctuations (see Fig.~S3b in SI~\cite{supmat}) as a function the inertial number $I$. The black solid line corresponds to $0.26\ d/(l_\textrm{c}\overline{\dot{\gamma}})$, with $l_\textrm{c}/d = a (1 + b/{I^{0.52}})$ with $a$=0.25 and $b$=1.5 found in panel a.} 
\label{Fig2}
\end{center}
\end{figure} 
A dense granular flow is characterized by rapid collisions involving sudden changes of the velocity direction and renewal of the contact network. Assuming that all these events occur at high frequency compared to the evolution of mean-field quantities, they can be described through a granular temperature~\cite{lois2005}. A reasonable assumption is to consider that this temperature is related to the local velocity fluctuations, through the proportionality relation $k_{\textrm{B}}T(z)\sim m\delta v^2(z)$, with $\delta v^2(z)= \langle| {v}(z,t)-\langle {v}(z,t)\rangle|^2\rangle$ the local variance of the velocity field  ${v}(z,t)$ along the flow direction. Figure~\ref{Fig2}b shows the evolution of the dimensionless standard deviation $\overline{\delta v} / (d \overline{\dot{\gamma}})$, where $\overline{A}=\frac 1H \int_0^H\textrm{d}z\,A(z)$ represents the thickness average of $A(z)$, as a function of the inertial number, for frictional and frictionless grains. We stress that the dimensionless standard deviation is independent of $z$ due to the Bagnold profile satisfied by $\langle v(z,t)\rangle$ (see Fig.~S2a in SI~\cite{supmat}) and the affine spatial behaviour of the variance observed in Fig.~\ref{Fig2}a. Interestingly, no matter the frictional nature of the grains, all the data reported here and in the literature collapse onto a single master curve showing a decrease of the relative velocity fluctuations with increasing inertial number. For small $I$ values ($I<0.07$), the dimensionless standard deviation decreases as $I^{-0.52\pm 0.01}$, while it seems to saturate to a constant value at large $I$~\cite{Kharel2017}. Interpolating the two asymptotic behaviours through a simple crossover form, one gets $\overline{\delta v}/(d \overline{\dot{\gamma}})=a (1 + b/I^{0.52})$, that fits well the data with $a=0.25$ and $b=1.5$ (see black line in Fig.~\ref{Fig2}a).

Let us now investigate the impact of the effective thermal energy on the grain dynamics. As shown by the time evolution of their thickness-averaged mean-square displacement along $z$ (see Fig. S3a in SI~\cite{supmat}), the grains globally diffuse perpendicularly to the flow direction, at long time with an associated thickness-averaged diffusion coefficient $\overline{D_z}$ increasing with the inclination angle $\theta$. Furthermore, as shown in Fig.~\ref{Fig2}c, $\overline{D_z}$ increases linearly with the thickness-averaged standard deviation $\overline{\delta v}$ of the velocity field. This linear relation can be understood from the thickness-averaged Kubo relation:
\begin{equation}
\overline{D_z} = \int \textrm{d}t\, \overline{\langle w(z,t) w(z,0)\rangle } \sim \overline{\tau(z)\delta v^2(z)}  \sim d\, \overline{\delta v}\ ,
\end{equation}
with $w(z,t)$ the velocity field along $z$, at position $z$ and time $t$, and where we assumed isotropic local velocity correlations of amplitude $\delta v^2(z)$ decaying in an exponential fashion over a local characteristic time $\tau(z)\sim d/\delta v(z)$. In addition, given the affine trends in Fig.~\ref{Fig2}b, one can show that $\overline{\tau}\sim d/\overline{\delta v}$. The thickness-averaged temporal correlations functions $\overline{\langle w(z,t) w(z,0)\rangle}$ of the velocity field along $z$, as calculated from the DEM trajectories (see Fig. S3b in SI~\cite{supmat}), appear to decay faster with increasing $\theta$. Neglecting long-time power-law tails, we can show that the exponential-decay time of $\overline{\langle w(z,t) w(z,0)\rangle}$ is well approximated by $\sim\overline{\tau}$. Besides, the velocity correlations suggest the existence of dynamic clusters that persist over the correlation time. We thus hypothesize the existence of a  characteristic, mesoscopic and \textit{a priori} $z$-dependent size $l_\textrm{c}(z)$ over which dynamic clusters persist during the time $\tau(z)$. This is reminiscent of the vortices discussed by Kharel and Rognon~\cite{Kharel2017}. As proposed by DeGiuli \textit{et al.}, these clusters produce an amplification of the velocity fluctuations that is estimated through a ``lever'' effect~\cite{DeGiuli2015,DeGiuli2016}. Specifically, one has
 $l_\textrm{c}(z)\sim d/[\tau(z) \dot{\gamma}(z)]$, and, with the definition $\tau \sim d/\delta v$, one gets $\delta v(z)/[d \dot{\gamma}(z)]\sim l_\textrm{c}(z)/d$ where the amplification factor appears clearly. Interestingly, since the left-hand side of the latter relation is independent of $z$, as discussed above, one gets that the dynamic-cluster size $l_{\textrm{c}}$ is in fact independent of $z$ for the inclined-plane configuration.
Figure~\ref{Fig2}d shows $\overline{\tau}$, as estimated from the thickness-averaged temporal correlations functions (see Fig. S3b in SI~\cite{supmat}), as a function of $I$. The data are in agreement with the relation $\overline{\tau}\sim d/[l_\textrm{c}\overline{\dot{\gamma}}]$ with the $l_\textrm{c}$ derived from the crossover expression between the two asymptotic regimes of Fig.~\ref{Fig2}a.
Since the expression for $l_\textrm{c}$  is independent of the frictional nature of the grains, this agreement suggests that the size of the dynamic clusters is mainly determined by the collisions between grains, but not by the microscopic friction between them. Furthermore, from Fig.~\ref{Fig2}a this dynamic-cluster size is expected to diverge at kinetic arrest -- which is reminiscent of the hypothetical cooperative length associated with the glass and jamming transitions. It should however be noted that some influence of the microscopic friction coefficient has been observed by DeGiuli and Wyart~\cite{DeGiuli2015,DeGiuli2016}, but for very small $I$ values that are well below the range accessed here.

In the following, we aim deriving the macroscopic rheological laws from the microscopic fluctuations and correlations. From dimensional analysis, we have recalled that a single parameter determines the flow properties. In the inclined-plane geometry, all dimensionless parameters are uniquely determined by the inclination angle $\theta\simeq \mu$. Therefore, the dimensionless ratio $Pd^3/(k_{\textrm{B}}T)$ should be constant in the layer for inclined-plane experiments. In a continuous mean-field approximation, the pressure field is hydrostatic, \textit{i.e.} $P(z)= \phi\rho_\textrm{p} g (H-z) \cos \theta \simeq \phi\rho_\textrm{p} g (H-z)$. It thus follows that the effective temperature must vary with the depth according to $T(z)\propto(H-z)$. As observed in Fig.~\ref{Fig2}b, apart from slight boundary deviations, the affine relation $\delta v^2(z)\propto(H-z)$ is satisfied for all the tested inclination angles $\theta$, which supports the definition of the effective temperature through $k_\textrm{B}T(z)\sim m \delta v^2(z)$. Interestingly, the effective temperature and the associated mechanical noise are maximal near the substrate and vanish at the free interface. This suggests that the collisions between mobile grains and the glued ones at the substrate is the source of temperature in the system. 
Furthermore, using the definition of the inertial number, the pressure can be written as $P(z)\sim {m \dot{\gamma}(z)^2}/{(dI^2)}$. Combining this relation with $\delta v(z) \sim l_\textrm{c} \dot{\gamma}(z)$, and the definition of the effective temperature, one gets $Pd^3/ (k_{\textrm{B}}T) \sim d^2/(l_\textrm{c}^2 I^2)$. 
The cluster size can be derived from free volume and cluster fractal shape arguments. Indeed, the required free volume to allow the motion of a grain implies the collective motion of $N_\mathrm{c}$ grains forming a dynamic cluster. The number of grains involved scale as $N_\textrm{c}\sim 1/(\phi_\textrm{c}-\phi)$. Assuming chain-like clusters with random walk-like geometry~\cite{karmakar2014,salez2015}, their size should be given by $l_\textrm{c}\propto N_\textrm{c}^{\nu}$ with $\nu\simeq 0.5-0.6$. 
The size of the cluster then scales with the packing fraction as $l_\textrm{c}\sim d/(\phi_\textrm{c}-\phi)^{\nu}$. 
Inserting this relation in the expression for the pressure yields $Pd^3/ (k_{\textrm{B}}T) \sim (\phi_\textrm{c}-\phi)^{2\nu}/I^2$. By identifying the latter relation with the equation of state (EOS) for hard-sphere fluids near the jamming transition~\cite{Torquato2010,Parisi2010}, \textit{i.e.} $Pd^3 / (k_\mathrm{B} T) = \phi_\mathrm{J}/(\phi_\mathrm{J}-\phi)$, one gets the dilatancy law:
\begin{equation}
(\phi_\textrm{c} - \phi) \sim I^{2/(2\nu+1)}\ . 
\end{equation}
with $0.91 \le 2/(2\nu+1) \le 1$, provided that we assume that $\phi_\mathrm{J}=\phi_\mathrm{c}$.
These dependencies in inertial numbers are in agreement with the observations. For the dilatancy law, Fig.~1d shows that the exponent $\alpha=2/(2\nu+1)$ is equal to $0.89\pm 0.10$ for frictional and $0.73\pm 0.34$ for frictionless grains. The large uncertainty observed for frictionless data is related to the lack of values at large $I$. For the cluster size,  the theory predicts a law $l_\textrm{c}\sim d/I^{\beta}$ with $\beta=2\nu/(2\nu+1)$. As shown in Fig.~2a, we observe $\beta=0.52\pm 0.01$ in very good agreement with the prediction for this exponent, \textit{i.e.} $0.5\le \beta \le0.54$. 
The universal agreement for both frictionless and frictional grains can be related to the evolution of the cluster size with inertial number, and reflects once again the dominance of collisions over friction in the dynamics. The validity of the hard-sphere-fluid EOS is probably limited to moderate inertial numbers, \textit{i.e.} $I\lesssim 0.5$, where the granular system can be considered as a fluid and where the mechanical noise ensures that no long-range correlations develop.

Let us finally propose a microscopic picture for the $\mu(I)$ rheological law. To do so, we consider the steady-state balance of driving and dissipated powers for a test grain located in a slab of thickness $d$ at height $z$. First, to estimate the driving contribution, we consider that the grain experiences the sum of gravitational and friction forces projected in the flow direction, and that $\theta$ and $\theta_\textrm{c}$ are small, leading to an effective driving force $\sim\rho_{\textrm{p}}\phi g (H-z) d^2 (\theta-\theta_\textrm{c})$. Since the grain moves over a distance $d$ within a time $\dot{\gamma}(z)^{-1}$, the net local driving power is $\dot{W}_\textrm{d}(z)\sim \rho_{\textrm{p}}\phi  g (H-z) d^3 (\theta-\theta_\textrm{c}) \dot{\gamma}(z)$. Secondly, we assume that the energy is mainly dissipated through the collisions with other grains, characterized by the characteristic decay time $\tau(z) \sim d/[l_\textrm{c} \dot{\gamma}(z)]$. The local power dissipated by collisions can thus be estimated by $\dot{W}_\textrm{c}(z) \sim m\delta v^2(z) /\tau(z)$. Balancing $\dot{W}_\textrm{d}(z)$ with $\dot{W}_\textrm{c}(z)$, and recalling that $\delta v\sim\dot{\gamma} l_\textrm{c}$ leads to $\dot{\gamma}^2 \sim g d  \phi (H-z)(\theta-\theta_\textrm{c})/l_\textrm{c}^{\,3}$. 
At small angles, and thus small $I$, Fig.~2a shows that the cluster size is adequately described by the relation: $l_\mathrm{c}\sim d \, I^{-\beta}$. Inserting this expression in the previous one, together with the definition of $I$, yields the general relation:
\begin{equation}
\dot{\gamma} \sim \left[{g\phi(H-z) \over d^2}\right]^{1/2} (\theta-\theta_\mathrm{c})^{1/(2-3\beta)}\ .
\label{equ_rheo}
\end{equation}
First, this expression is compatible with the $z$-dependency of the Bagnold velocity profile, $\langle v(z,t)-v(0,t)\rangle \propto H^{3/2} - (H-z)^{3/2}$. Secondly, recalling that $\mu\simeq\theta$, as well as the definition of $I$, Eq.~(\ref{equ_rheo}) yields the friction law $\mu -\mu_\textrm{c}\sim I^{2-3\beta}$. Considering the theoretical range of $\beta$ values, $0.5\le \beta \le0.54$, we obtain a prediction for the exponent, \textit{i.e.} $0.38 \le (2-3\beta)\le 0.5$, in close agreement with the law $\mu-\mu_c \sim I^{0.40\pm0.01}$ observed for frictionless grains shown in Fig.~\ref{Fig1}c.


We emphasize that the proposed model, based on a fractal dimension for the chain-like clusters related to a simple random walk, is able to properly predict three different laws based on the measurements of velocity fluctuations ($\delta v/\dot{\gamma} \sim l_c \sim d I^{-0.52}$), packing fractions ($\phi_c-\phi\sim I^{0.9}$) and flow velocity ($\mu-\mu_c\sim I^{0.4}$). 

One may naively expect Eq.~(\ref{equ_rheo}) to also hold for frictional systems, since the velocity fluctuations and cluster size behave similarly with the inertial number for both frictional and frictionless systems. However, it can not explain the $\mu -\mu_\textrm{c}\sim I$ relation observed for frictional grains in Fig.~\ref{Fig1}c. This disagreement is in fact not surprising. In the derivation of Eq.~(\ref{equ_rheo}), it is assumed that all the energy dissipation arises from collisions between grains. This is a very reasonable assumption for frictionless systems, but an additional source of dissipation is expected from the mobilization of frictional contacts. Unfortunately, including frictional dissipation in a theoretical model for dense granular flows remains a highly debated issue~\cite{Azema2014,DeGiuli2015,DeGiuli2016}. Nevertheless, interestingly, Fig.~\ref{Fig1}c shows that for large-enough inertial numbers, the data obtained for frictional systems collapse onto the law of frictionless systems. This observation suggests that, in the limit of large $I$, the energy dissipation is universal and of collisional origin.
 
In summary, from numerical simulations and inspection of the literature data, we show that the dilatancy law is identical for frictionless and frictional assemblies. This law can be further rationalized from a comparison between: i) the equation of state constructed from the hydrostatic pressure, an effective granular temperature related to velocity fluctuations, as well as the inertial number; and ii) the equation of state of hard-sphere fluids near the jamming transition. In contrast, the macroscopic friction laws are observed to differ for frictionless and frictional assemblies. In the former case, we can rationalize the observations from a power balance at the grain level, involving gravity, effective friction, and collisions. We recover as well the Bagnold profile for the  local average velocity field. The derivation of a macroscopic friction law for frictional assemblies remains an open question and should involve an additional dissipation term related to the formation of frictional contacts.

The authors thank Yoel Forterre, Olivier Pouliquen, Olivier Dauchot and Pierre Soulard for enlightening suggestions. This work benefited from financial support of the Fonds National de la Recherche Scientifique through the PDR research project T.0251.20 ``Active matter in harmonic traps", the Foundation for Training in Industrial and Agricultural Research, the Agence Nationale de la Recherche (ANR-21-ERCC-0010-01 \textit{EMetBrown}, ANR-21-CE06-0029 \textit{SOFTER}, ANR-21-CE06-0039 \textit{FRICOLAS}), and the French Friends of the Hebrew University of Jerusalem and the Scopus Foundation. The authors also thank the Soft Matter Collaborative Research Unit, Frontier Research Center for Advanced Material and Life Science, Faculty of Advanced Life Science at Hokkaido University, Sapporo, Japan.

\end{document}


\title{Microscopic foundation of the $\mu$(I) rheology for dense granular flows on inclined planes \\ -- Supplementary Information --}
\author{Denis Dumont}
\affiliation{Laboratoire Interfaces $\&$ Fluides Complexes, Universit\'e de Mons, 20 Place du Parc, B-7000 Mons, Belgium.}
\author{Haggai Bonneau}
\affiliation{UMR CNRS Gulliver 7083, ESPCI Paris, PSL Research University, 75005 Paris, France.}
\author{Thomas Salez}
\affiliation{Univ. Bordeaux, CNRS, LOMA, UMR 5798, F-33400 Talence, France.}
\author{Elie Raphael}
\affiliation{UMR CNRS Gulliver 7083, ESPCI Paris, PSL Research University, 75005 Paris, France.}
\author{Pascal Damman}
\email{pascal.damman@umons.ac.be}
\affiliation{Laboratoire Interfaces $\&$ Fluides Complexes, Universit\'e de Mons, 20 Place du Parc, B-7000 Mons, Belgium.}
\date{\today}
\maketitle

\section{Friction}

\begin{figure}[h]
\begin{center}
\includegraphics[width=0.49\linewidth]{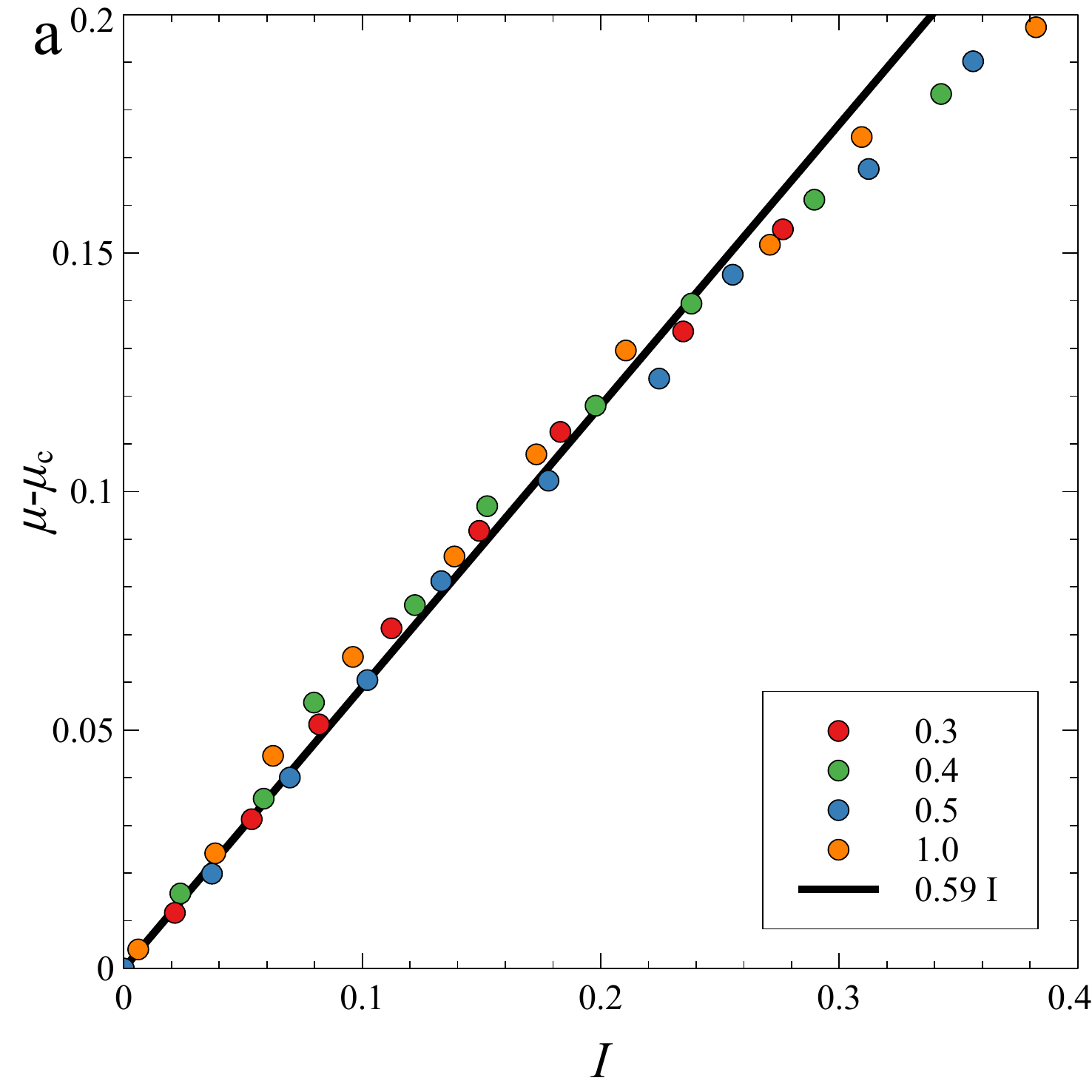}
\includegraphics[width=0.49\linewidth]{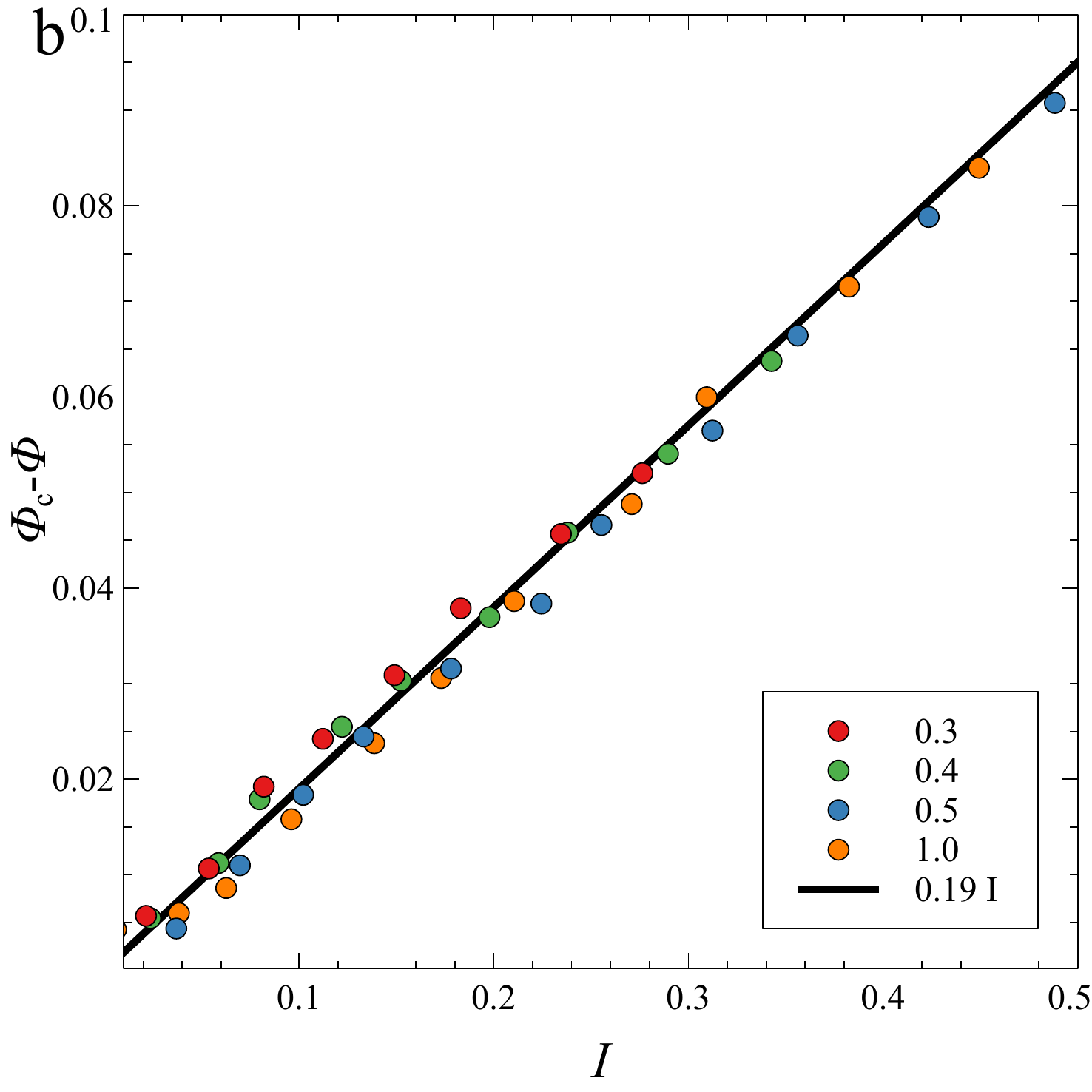}
\caption{a) Shifted effective friction coefficient, $\mu- \mu_{\textrm{c}}$ as a function of the inertial number $I=\dot{\gamma}d \sqrt{\rho_{\textrm{p}}/P}$, obtained for a layer of grains with an initial layer thicknesses $H_{\textrm{i}}=20d$ and various microscopic sliding frictions $\mu_s$ as indicated. The solid line indicates a linear fit of the small $I$ values as provided in legend. b) Shifted volume fraction $\phi_\textrm{c}-\phi$ as a function of the inertial number $I$. The solid line indicates a linear fit as provided in legend.}
\label{Fig1s}
\end{center}
\end{figure}

We carry out additional simulations and change the value of the microscopic sliding friction coefficient $\mu_\textrm{s}$, from 0.3 to 1. As shown in Fig~\ref{Fig1s}, we note that it only shifts the threshold values $\mu_\textrm{c}(\mu_\textrm{s})$ and $\phi_\textrm{c}(\mu_\textrm{s})$ but has no impact on the scaling laws. By simply subtracting the threshold value, we obtain a collapse of all our data on a single master curve. 

\section{Bagnold-like flows}

\begin{figure}[h]
\begin{center}
\includegraphics[width=1\linewidth]{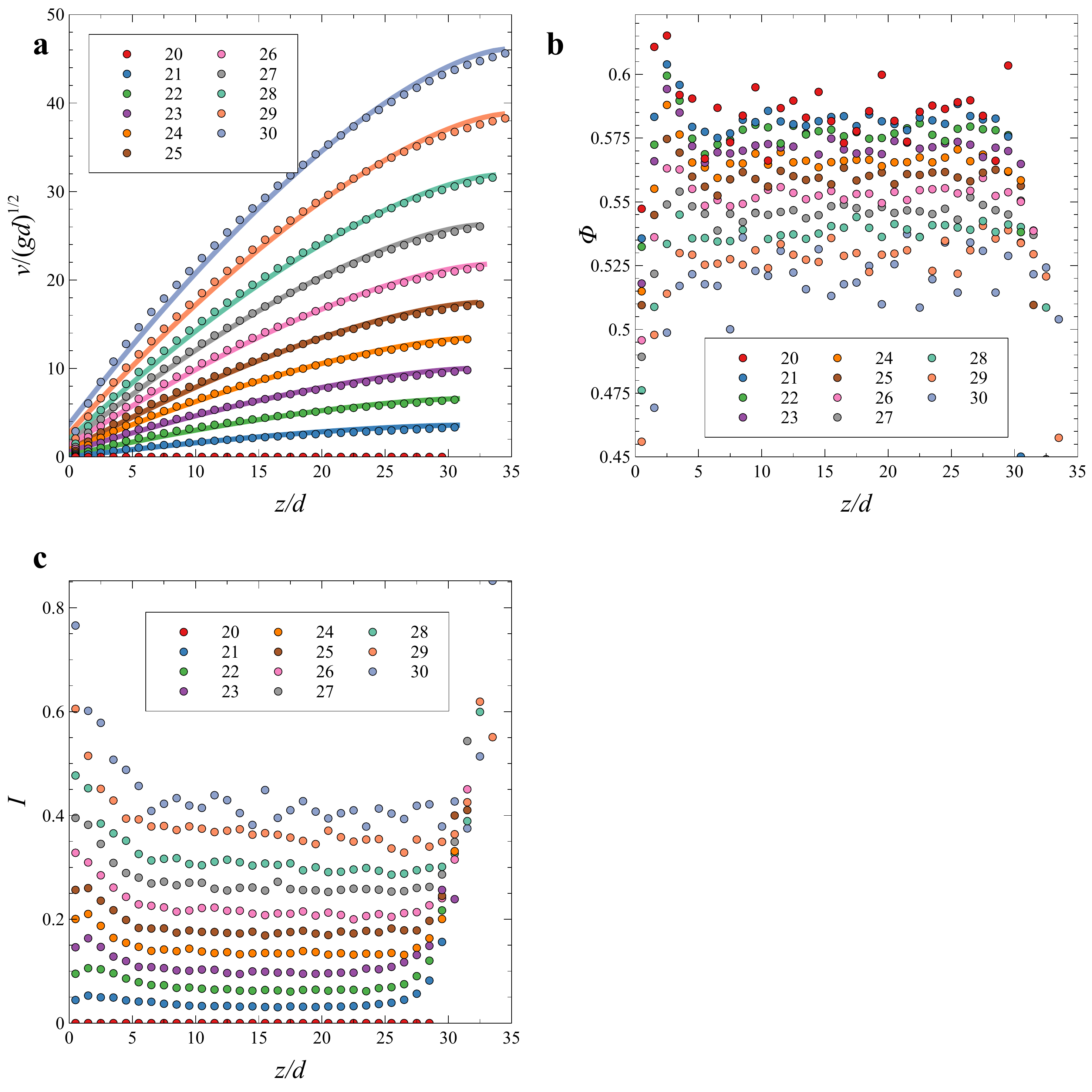}
\caption{Evolution of the rescaled local velocity (a), volume fraction $\phi$ (b) and inertial number $I$ (c) as functions of rescaled normal coordinate $z/d$ for a layer of frictional grains initially characterized by $H_{\textrm{i}}=30d$ and various inclination angles $\theta$ as indicated. The curves in (a) corresponds to the best fit of the Bagnold profile expression.}
\label{Fig2s}
\end{center}
\end{figure}

In order to check the validity of our DEM simulations of granular flows on inclined planes, we study in details the properties of the flow. We observe that the local time-averaged velocity profile is well described by a Bagnold profile $\langle v(z)-v(0)\rangle \sim H^{3/2}-(H-z)^{3/2}$ (see Fig.~\ref{Fig2s}a). Besides, the inertial number $I$ (see Fig.~\ref{Fig2s}b) and the volume fraction $\phi$ (see Fig.~\ref{Fig2s}c) remain mostly constant throughout the layer (except at both boundaries) for all the studied inclination angles. These observations are in good agreement with previously reported observations~\cite{GDRMiDi2004,Baran2006,Silbert2001,Andreotti2013,Silbert2003}.

\section{Velocity correlations and diffusion}

\begin{figure}[h]
\begin{center}
\includegraphics[width=1\linewidth]{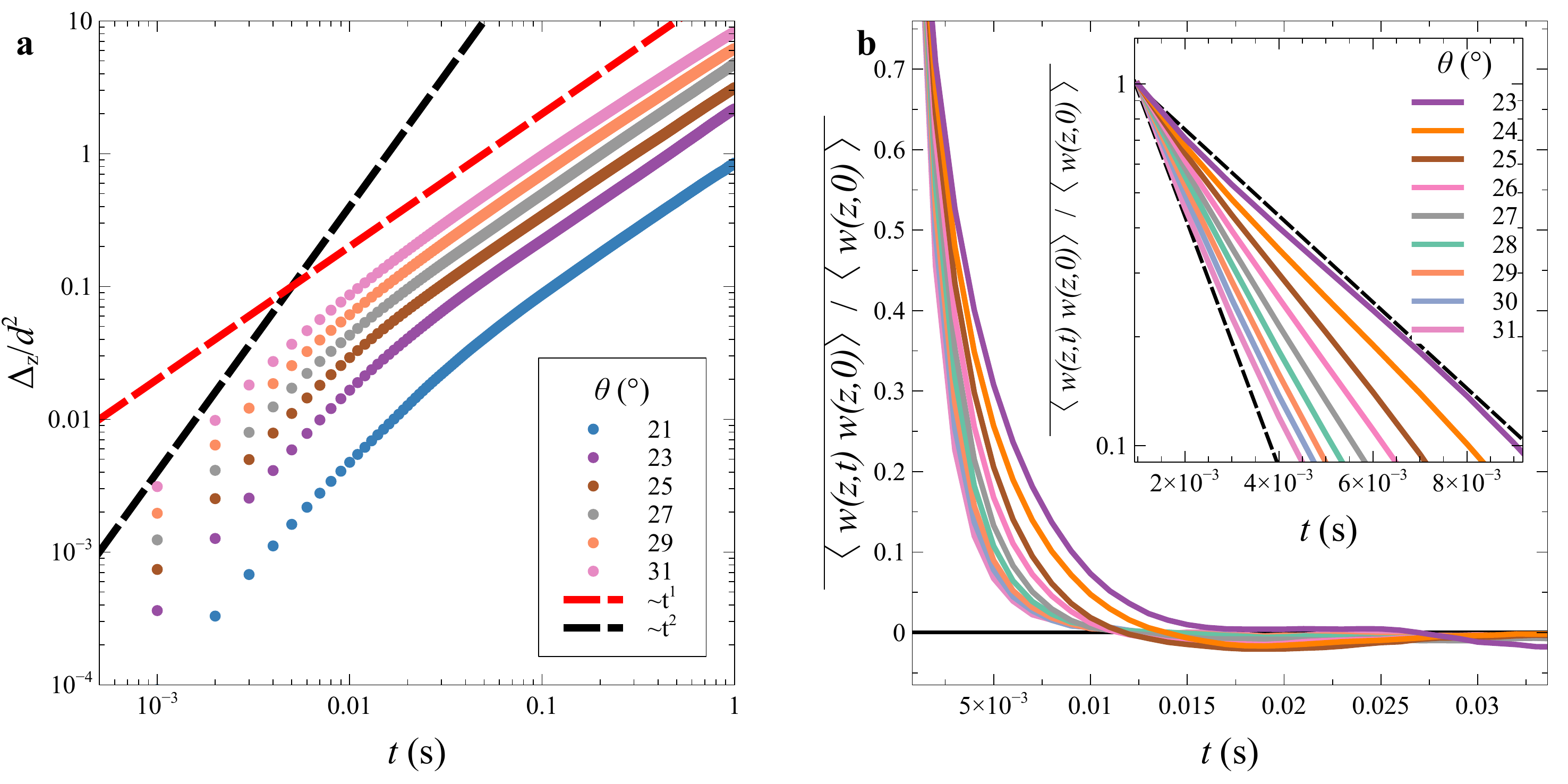}
\caption{(a) Mean square displacement $\Delta_\textrm{z}$ along the $z$ coordinate averaged over all the grains and (b) time correlations of the velocity following $z$ for a layer of frictional grains initially characterized by $H_{\textrm{i}}=30d$ and various inclination angles as indicated. Inset: Log-lin plot of the velocity correlations.}
\label{Fig3s}
\end{center}
\end{figure}

As shown in Fig.~\ref{Fig3s}a, we calculate the mean square displacement ($\Delta_\textrm{z}$, MSD) along the $z$ coordinate averaged over all the grains for various inclination angles. From those MSD, the thickness-averaged diffusion coefficient, $\overline{D_\textrm{z}}$ is evaluated by fitting the data in the diffusive regime, \textit{i.e. } where $\Delta_\textrm{z}=2 \overline{D_z} t$. In addition, we also evaluate the time correlation of the $z$-component of the velocity $w$, averaged over the layer thickness, for various inclination angles. The thickness averaged correlation times, $\overline{\tau}$ were obtained by fitting the correlation curves with decreasing exponentials of the type $\sim e^{-t/\overline{\tau}}$. 

